\documentclass[amsmath,floatfix,twocolumn,superscriptaddress]{revtex4-1}
\usepackage{subfigure}
\usepackage{siunitx}
\usepackage{amssymb}
\usepackage{amsmath}
\usepackage{graphicx}
\usepackage{array}
\usepackage{dcolumn}
\usepackage{psfrag}
\usepackage{bm}
\usepackage{color}
\usepackage{multirow}

\begin{document}

\title{Surface Physics of Low-Dimensional Pnictogen Chalcohalides and Its Impact on \\ 
                        Photovoltaic Technologies}

\author{Cibrán López}
\email{cibran.lopez@upc.edu}
    \affiliation{Departament de Física, Universitat Politècnica de Catalunya, 08034 Barcelona, Spain}
    \affiliation{Research Center in Multiscale Science and Engineering, Universitat Politècnica de Catalunya,
                 Campus Diagonal-Besòs, Av. Eduard Maristany 10–14, 08019 Barcelona, Spain}

\author{David Rovira}
    \affiliation{Research Center in Multiscale Science and Engineering, Universitat Politècnica de Catalunya,
                 Campus Diagonal-Besòs, Av. Eduard Maristany 10–14, 08019 Barcelona, Spain} 
    \affiliation{Department of Electronic Engineering, Universitat Politècnica de Catalunya, 08034 Barcelona, Spain}

\author{Álex Jiménez-Arguijo}
   \affiliation{Research Center in Multiscale Science and Engineering, Universitat Politècnica de Catalunya,
                 Campus Diagonal-Besòs, Av. Eduard Maristany 10–14, 08019 Barcelona, Spain}
    \affiliation{Department of Electronic Engineering, Universitat Politècnica de Catalunya, 08034 Barcelona, Spain}

\author{Zacharie Jehl}
    \affiliation{Research Center in Multiscale Science and Engineering, Universitat Politècnica de Catalunya,
                 Campus Diagonal-Besòs, Av. Eduard Maristany 10–14, 08019 Barcelona, Spain}
    \affiliation{Department of Electronic Engineering, Universitat Politècnica de Catalunya, 08034 Barcelona, Spain}

\author{Edgardo Saucedo}
    \affiliation{Research Center in Multiscale Science and Engineering, Universitat Politècnica de Catalunya,
                 Campus Diagonal-Besòs, Av. Eduard Maristany 10–14, 08019 Barcelona, Spain}
    \affiliation{Department of Electronic Engineering, Universitat Politècnica de Catalunya, 08034 Barcelona, Spain}

\author{Claudio Cazorla}
\email{claudio.cazorla@upc.edu}
    \affiliation{Departament de Física, Universitat Politècnica de Catalunya, 08034 Barcelona, Spain}
    \affiliation{Research Center in Multiscale Science and Engineering, Universitat Politècnica de Catalunya,
                 Campus Diagonal-Besòs, Av. Eduard Maristany 10–14, 08019 Barcelona, Spain}
    \affiliation{Institució Catalana de Recerca i Estudis Avançats (ICREA), Passeig Lluís Companys 23, 08010 Barcelona, Spain}

\begin{abstract}
\textbf{Abstract.}~Low-dimensional pnictogen chalcohalides (MChX; M = Bi, Sb; Ch = S, Se; X = I, Br) are promising non-toxic 
semiconductors for solar energy conversion, yet the power conversion efficiencies of MChX solar cells remain below 10\%, far 
from their Shockley-Queisser limit of $\sim$30\%. Efficient charge extraction depends critically on the band-edge alignment 
between the absorber and the charge-selective contacts, which is a surface property. Here, we combine first-principles calculations 
with device-level modelling to investigate the surface energetics and band alignments across the MChX family. An exhaustive 
sampling of crystal orientations, including low-symmetry facets, identifies the (011) and (010) surfaces as the most stable 
terminations in all eight compounds. Their formation energies differ by only $\approx 0.01$--$0.03~\mathrm{J\,m^{-2}}$, yet 
their band edges are shifted rigidly with respect to each other by up to $0.8$~eV. Drift--diffusion simulations of BiSBr show 
that the coexistence of such facets can reduce the open-circuit voltage by up to $0.6$~V, even in an otherwise ideal, defect-free 
absorber. Surface orientation thus emerges as a hidden design parameter for charge extraction. This study may help to explain 
the gap between the theoretical and experimental efficiencies of MChX solar cells, and it provides new design principles for 
MChX-based photovoltaic and photocatalytic technologies.
\\

\end{abstract}

\maketitle

\section{Introduction}
\label{sec:intro}
\begin{figure*}[t]
    \centering
        \includegraphics[width=1.0\textwidth]{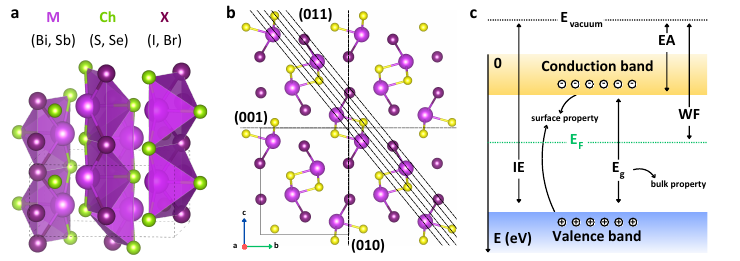}
        \caption{\textbf{MChX atomic structure, surfaces and band levels.}
        \textbf{a.}~Orthorhombic ($Pnma$ space group) crystal structure of bulk MChX.
        \textbf{b.}~The (010) and (001) planes are shown as representative surface orientations. The (011) orientation is 
        also illustrated together with non-equivalent terminations (dashed lines). 
        \textbf{c.}~Schematic band diagram showing the valence band maximum (VBM) and conduction band minimum (CBM) relative 
        to vacuum. While the band gap, $E_{g}$, is an intrinsic bulk property, the VBM and CBM levels depend on surface 
        orientation and termination. WF stands for work function, $E_{\rm F}$ for electronic Fermi level, EA for electron affinity, 
        and IE for ionization energy.}
        \label{fig1}
\end{figure*}

The discovery of new semiconductors with tailored optoelectronic properties is central to the development of technologies for 
energy conversion, storage, and sensing. Pnictogen chalcohalides (MChX, where M = Bi, Sb; Ch = S, Se; and X = I, Br; Fig.~\ref{fig1}a) 
have recently emerged as a compelling family of non-toxic semiconductors, featuring band gaps in the range of $1.2$--$2.1$~eV and 
strong absorption coefficients of $25$--$66$~$\mu$m$^{-1}$ \cite{Lopez2024,Ganose2016,Nielsen2025}. They exhibit robust thermodynamic 
stability \cite{Nie2019,Li2024} and can be synthesized at relatively low temperatures (below $300^\circ$C) \cite{Cano2023}. Taken 
together, these properties make MChX compounds attractive candidates for scalable and environmentally responsible photovoltaic 
technologies.

Nevertheless, despite these favorable intrinsic properties, experimental power conversion efficiencies (PCE) of MChX-based
solar cells remain below $10$\%, far from their theoretical Shockley-Queisser limit of $\sim 30$\% \cite{Miliaieva2025}. Part 
of this discrepancy can be attributed to non-radiative recombination at point defects \cite{Lopez2025,Lopez2026}; however, 
device-architecture limitations are likely to be equally important. In particular, the relative band alignment between the 
light-absorbing material and the charge-selective contact layers is a crucial parameter for efficiently extracting photogenerated 
carriers and generating substantial photocurrents.

A photovoltaic device must extract electrons at one terminal and holes at the other, while blocking the opposite charge carrier
from reaching each contact. This selectivity is not an intrinsic property of a single material, rather, it emerges from the
relative alignment of the transport-layer electronic bands with respect to those of the light absorber. Specifically, under 
contact equilibrium, the conduction band minimum (CBM) of the electron transport layer should align closely with that of the 
absorber to allow facile electron flow, while its valence band maximum (VBM) should lie well below that of the absorber, creating 
a large barrier that blocks holes. At the hole transport layer, the situation is inverted: the VBM should align with that of the 
absorber, while the CBM should be shifted to higher energy to block electron flow. If either of these offsets is not satisfied, 
the corresponding contact layer loses its selectivity, allowing both carrier types to reach the electrode and opening an undesired 
direct recombination pathway.

Band alignment is fundamentally a surface property: it depends strongly on the surface orientation and termination. As a result,
different facets of the same material can potentially exhibit markedly different VBM and CBM levels. Consequently, for a given 
choice of absorber and charge-selective contacts, the magnitude of the photogenerated current in a solar cell can vary substantially 
depending on which surfaces are exposed at the interfaces. Identifying the energetically most favorable surfaces, together with 
their associated VBM and CBM levels, is therefore essential for connecting nanoscale surface energetics with macroscopic photovoltaic 
device performance.

Despite its practical relevance to solar cell engineering, the surface physics of MChX compounds has been largely
unexplored, and their facet competition and associated band alignments remain poorly understood. To the best of our
knowledge, only a few recent experimental and theoretical studies \cite{Cano2025,Rovira2025,Lynch2026} have addressed the
energetic and electrostatic properties of MChX surfaces, and these have done so with limited atomistic resolution and
without considering low-symmetry surfaces that may be energetically competitive. This lack of knowledge could be hindering 
the development of MChX-based solar cells with PCE approaching their theoretical upper limit of approximately 
$30$\% \cite{Miliaieva2025}.

In this work, we present a comprehensive theoretical investigation of the surface energetics and band-edge alignments across
the MChX family (M = Bi, Sb; Ch = S, Se; and X = I, Br) using state-of-the-art first-principles density functional theory (DFT) 
simulations. Our analysis reveals that the (010) and (011) surfaces are energetically the most favorable and nearly degenerate, 
yet they exhibit markedly different band-edge alignments (one another may differ by up to $0.8$~eV). Using optical 
solar-cell device modeling based on drift-diffusion simulations, we quantify the extent to which this diversity of band 
alignments between energetically competitive surfaces can affect photovoltaic device performance. Our results thus establish 
fundamental design principles for MChX-based solar cells and provide a generalizable framework applicable to other emerging 
solar-cell technologies.

\begin{figure*}[t]
    \centering
        \includegraphics[width=1.0\textwidth]{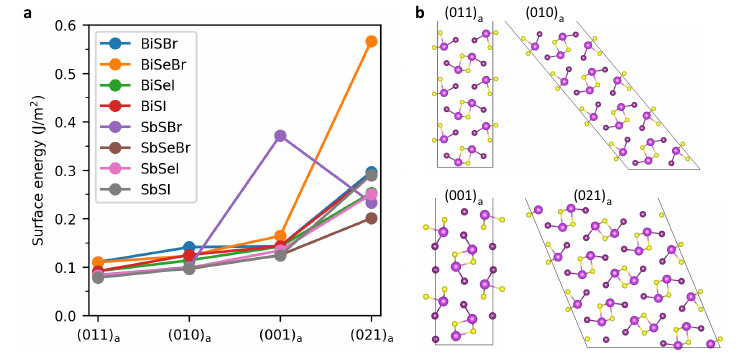}
        \caption{\textbf{Surface formation energies of MChX.}
         \textbf{a}~Surface formation energies of the four lowest-energy surface orientations for the eight MChX compounds 
          analysed in this study, with the $(011)$ and $(010)$ facets being the most stable. Lines are guides to the eye.
         \textbf{b}~Ball-stick representation of the relaxed atomic structures of the corresponding surface slabs.
         The ``a'' subscript indicates a specific surface termination that can change from one compound to the other 
         (see Supplementary Figs.~1--8 for accessing that information). }
        \label{fig2}
\end{figure*}

\section{Results}
\label{sec:results}
Bulk MChX semiconductors crystallize in an orthorhombic phase (space group $Pnma$) consisting of quasi-one-dimensional columns
weakly coupled by van der Waals interactions (Fig.~\ref{fig1}a) \cite{Cano2025}. Our DFT geometry optimizations (Methods) yield 
lattice parameters in excellent agreement with the reported experimental values \cite{Lopez2024}, and both simulations and 
experiments confirm that MChX compounds are thermodynamically stable against decomposition into competing secondary phases 
\cite{Cano2023,Li2024,Lopez2025,Lopez2026}. In the following, we first present and discuss the surface formation energies and
band-edge alignments obtained for eight exemplary MChX compounds. We then quantify, by means of macroscopic optoelectronic 
modeling, the possible impact of these atomistic results on the conversion performance of photovoltaic devices.

\subsection{Surface formation energy}
\label{subsec:energy}
A crystal can, in principle, expose an infinite number of facets, each defined by its Miller indices $(h,k,l)$ and its possible
atomic terminations (Fig.~\ref{fig1}b). The number of inequivalent surfaces grows rapidly with increasing Miller indices, since
high-$(h,k,l)$ surfaces entail low-symmetry terminations with a large number of dangling bonds and undercoordinated atoms. It is
therefore generally assumed that surfaces with higher Miller indices exhibit higher surface formation energies ($E_{\rm surface}$, 
Methods) than high-symmetry low-$(h,k,l)$ surfaces \cite{Tamilselvan2016}. In practice, this premise translates into an expected 
prevalent formation of high-symmetry low-$(h,k,l)$ surfaces during materials synthesis \cite{Herring1951}.

This assumption, however, is not universally valid, and its physical basis deserves scrutiny. The magnitude of the Miller
indices $(h,k,l)$ is fundamentally a crystallographic convention, tied to the choice of a conventional (often non-primitive)
unit cell, rather than a direct physical measure of the number of broken bonds per unit surface area. What actually governs
the surface formation energy is the planar atomic density and the strength of the bonds severed upon cleavage, neither of
which scales monotonically with increasing Miller indices \cite{Sun2018}. Moreover, in structurally anisotropic materials
with strongly directional bonding, such as MChX, a nominally high-index plane can run parallel to natural cleavage or
high-density atomic planes, thereby cutting through fewer bonds than several low-index terminations \cite{Sun2018,Siddique2010}.
Beyond packing density, surface relaxation and reconstruction offer a further mechanism by which high-index surfaces can fall
below their naively expected energy, as the rearrangement of undercoordinated atoms can partially or fully saturate dangling
bonds that would otherwise dominate the unrelaxed surface energy. Surface stability should therefore be assessed through
explicit energy calculations across an exhaustive set of orientations, rather than inferred from Miller index magnitude alone.

To perform extensive and systematic explorations of the formation energy of MChX surfaces, we used our in-house-developed \verb!SurfAIcing! 
software \cite{surfaicing1,surfaicing2}. By relying on foundational machine-learned interatomic potentials (ML-IAPs) \cite{mace} 
and the Python library \verb!pymatgen! \cite{Tran2016}, \verb!SurfAIcing! can efficiently generate and relax surfaces, and automatically
postprocess all the required DFT data to produce surface formation energy rankings and other related properties (e.g., band-edge
alignments, Fig.~\ref{fig1}c). Specifically, we first generated all possible MChX surfaces and surface terminations with Miller 
indices $h,k,l \leq 2$ and relaxed them with ML-IAPs \cite{mace}. Subsequently, we fully relaxed the ML-IAP-generated structures 
using DFT methods (Methods) and calculated their surface formation energies (Supplementary Figs.~1--8). In summary, the four 
lowest-energy surfaces of each of the eight MChX compounds considered in this study are shown in Fig.~\ref{fig2}.

\begin{figure*}[t]
    \centering
        \includegraphics[width=1.0\textwidth]{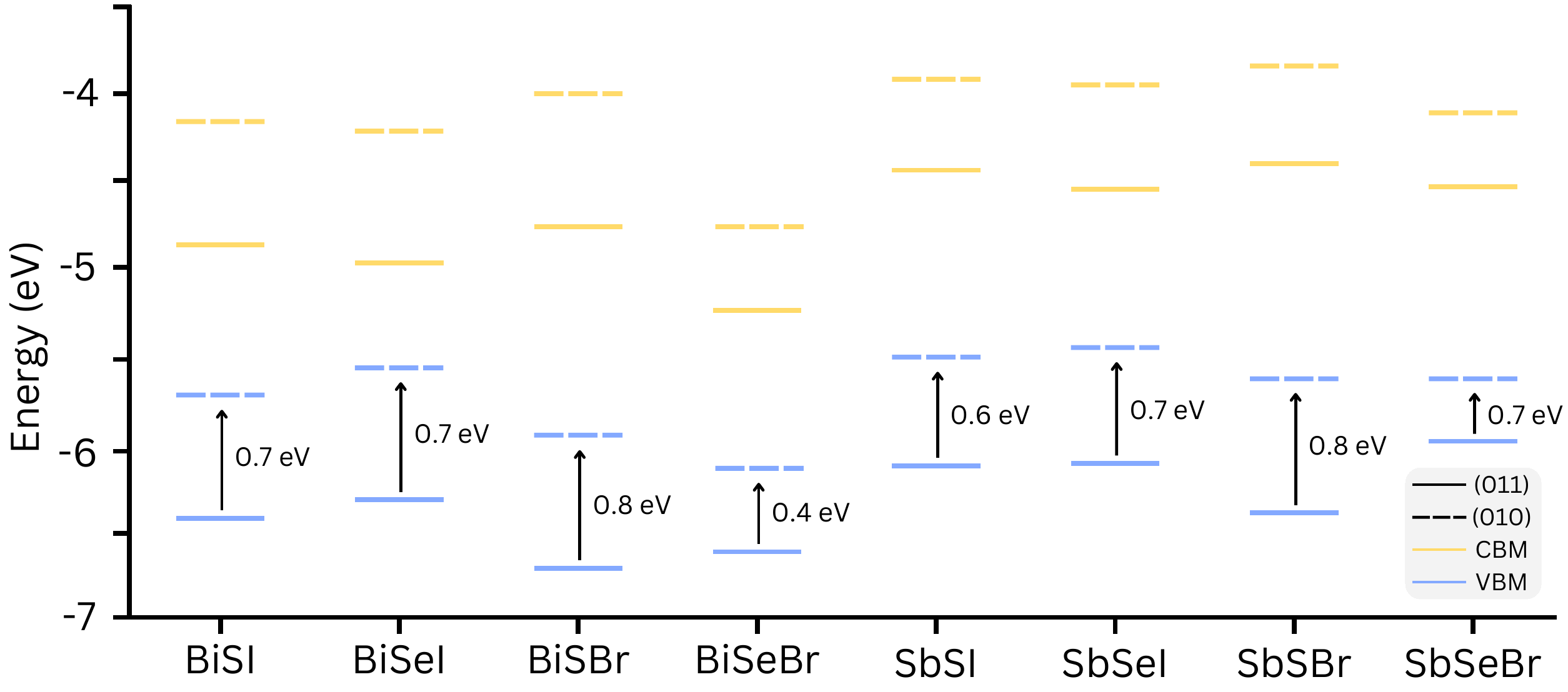}
        \caption{\textbf{Band alignments of MChX.}
        Valence band maximum (blue) and conduction band minimum (yellow) relative to the vacuum level for the $(011)$ (solid lines) 
        and $(010)$ (dashed lines) surfaces of the eight analysed MChX compounds. Arrows and numbers indicate the VBM shift between 
        the two facets.}
        \label{fig3}
\end{figure*}

Figure~\ref{fig2}a shows that the surface formation energies of the eight MChX compounds span a wide range, from $\approx 0.08$ 
to $\approx 0.57~\mathrm{J\,m^{-2}}$, and that the four lowest-energy orientations follow a remarkably consistent hierarchy 
across the family. In all cases, the $[011]$ orientation is the most stable, although the specific surface termination may vary
from one material to the other (Supplementary Figs.~1--8), with $E_\mathrm{surface}^{(011)}$ ranging from 
$0.078~\mathrm{J\,m^{-2}}$ (SbSI) to $0.110~\mathrm{J\,m^{-2}}$ (BiSBr and BiSeBr) (Table~1). Likewise, $(010)$ is the second most 
stable facet in all cases, with $E_\mathrm{surface}^{(010)}$ between $0.096~\mathrm{J\,m^{-2}}$ (SbSeBr) and 
$0.141~\mathrm{J\,m^{-2}}$ (BiSBr) (Table~1). The energy difference between these two facets is small, only 
$0.013$--$0.034~\mathrm{J\,m^{-2}}$, whereas the $(001)$ and $(021)$ surfaces lie systematically higher in energy. The $(001)$ 
values cluster between $\approx 0.12$ and $0.17~\mathrm{J\,m^{-2}}$, and the $(021)$ values between $\approx 0.20$ and 
$0.30~\mathrm{J\,m^{-2}}$, i.e., roughly $2.5$--$3.5$ times those of $(011)$. 

The absolute values of $E_\mathrm{surface}$ are low, of the order of $0.1~\mathrm{J\,m^{-2}}$ for the most stable facets, in 
comparison to those of, for instance, binary oxides \cite{binaryoxide1,binaryoxide2}, as expected for quasi-one-dimensional 
compounds in which neighbouring columns interact only weakly through van der Waals forces. Importantly, the ordering shown in 
Fig.~\ref{fig2}a does not follow the magnitude of the Miller indices: the $(011)$ facet is more stable than the lower-index 
$(010)$ and $(001)$ facets in every compound. This observed trend confirms that the Miller-index magnitude is a poor predictor 
of surface stability in MChX, and that low-symmetry orientations can be energetically favoured over high-symmetry ones.

Systematic chemical trends are also apparent. Sb-based compounds display lower surface energies than their Bi-based counterparts 
for both leading facets, by roughly $20\%$ on average (e.g., $E_\mathrm{surface}^{(011)} = 0.078$ and $0.091~\mathrm{J\,m^{-2}}$ 
for SbSI and BiSI, respectively), which is likely related to differences in bonding between the two pnictogens. In Bi-based 
compounds, replacing I by Br also raises the energy of both leading facets (e.g., $E_\mathrm{surface}^{(010)} = 0.125$ and 
$0.141~\mathrm{J\,m^{-2}}$ for BiSI and BiSBr), whereas the effect is marginal in Sb-based compounds. Two data points deviate
markedly from the general behaviour: the $(001)$ surface of SbSBr ($\approx 0.37~\mathrm{J\,m^{-2}}$) and the $(021)$ surface 
of BiSeBr ($\approx 0.57~\mathrm{J\,m^{-2}}$). We attribute these outliers to termination-specific effects, such as the severing 
of a large number of strong intra-column bonds or the absence of an effective surface reconstruction (Fig.~\ref{fig2}b). Neither 
affects the stability ranking, since both occur in facets that are already disfavoured. Overall, the $(011)$ and $(010)$ surfaces 
are the only two facets that are energetically competitive in all eight compounds, and their small energy difference suggests 
that both can be expressed depending on the synthesis conditions. As discussed next, this is decisive because these two facets 
exhibit markedly different band-edge alignments (Fig.~\ref{fig3} and Table~1).

\begin{table*}[t]
    \centering
    \begin{tabular}{ccccccccc}
        \hline
        \hline
        \quad Material \quad & \quad $E_{g}$ \quad & \quad  $E_{\rm surface}^{\rm (011)}$ \quad & \quad VBM$^{\rm (011)}$ \quad & 
        \quad CBM$^{\rm (011)}$ \quad & \quad $E_{\rm surface}^{\rm (010)}$ \quad & \quad VBM$^{\rm (010)}$ \quad & \quad
         CBM$^{\rm (010)}$ \quad & \quad $|\Delta V|$ \quad \\ 
               & (eV) & (J~m$^{-2}$) & (eV) & (eV) & (J~m$^{-2}$) & (eV) & (eV) & (eV) \\ 
        \hline
        BiSI   & 1.49 & 0.091 & -6.39 & -4.90 & 0.125 & -5.70 & -4.21 & 0.69 \\
        BiSeI  & 1.25 & 0.091 & -6.22 & -4.97 & 0.114 & -5.55 & -4.30 & 0.67 \\
        BiSBr  & 1.84 & 0.110 & -6.65 & -4.81 & 0.141 & -5.85 & -4.01 & 0.80 \\
        BiSeBr & 1.30 & 0.110 & -6.55 & -5.25 & 0.123 & -6.17 & -4.87 & 0.38 \\
        SbSI   & 1.70 & 0.078 & -6.12 & -4.42 & 0.097 & -5.49 & -3.79 & 0.63 \\
        SbSeI  & 1.50 & 0.083 & -6.11 & -4.61 & 0.098 & -5.45 & -3.95 & 0.66 \\
        SbSBr  & 1.93 & 0.083 & -6.40 & -4.47 & 0.099 & -5.61 & -3.68 & 0.79 \\
        SbSeBr & 1.45 & 0.083 & -5.99 & -4.54 & 0.096 & -5.32 & -3.87 & 0.67 \\
        \hline
        \hline
    \end{tabular}
    \caption{\textbf{Surface formation energies.} Calculated surface formation energies ($E_{\text{surface}}$) and valence band maximum 
    positions (VBM) for the two most stable MChX surface terminations. Results are shown for the (011) and (010) facets. 
    Band gaps ($E_{g}$) and VBM values are given in eV, while $E_{\text{surface}}$ is given in J/m$^{2}$. The comparison highlights the 
    nearly degenerate stability of (011) and (010) surfaces, despite their markedly different band edge alignments.}
    \label{table1}
\end{table*}

\subsection{Band alignments}
\label{subsec:bandalignments}
While the band gap is an intrinsic bulk property, determined primarily by the bulk geometry and chemical composition, the 
absolute alignment of the band edges with respect to the vacuum level is a surface property that is sensitive to atomic 
termination, reconstruction, and crystallographic orientation (Fig.~\ref{fig1}c). This distinction is particularly relevant 
for photovoltaic applications, where the band alignment between the light-absorbing layer and the charge-selective contacts 
dictates whether photogenerated carriers are efficiently extracted or lost to recombination (Sec.~\ref{sec:intro}). Accurate 
knowledge of how the band edges of the absorber depend on the exposed surface orientation is therefore essential for the 
design of efficient solar cells. In the case of MChX, we have seen in the previous section that the (011) and (010) surfaces 
are close in energy and may thus coexist in real samples, so it is necessary to assess how much their band edges differ. 
Fig.~\ref{fig3} and Table~1 summarise the VBM and CBM positions of both facets for the eight considered compounds (Methods).

\begin{figure*}[t]
    \centering
        \includegraphics[width=1.0\textwidth]{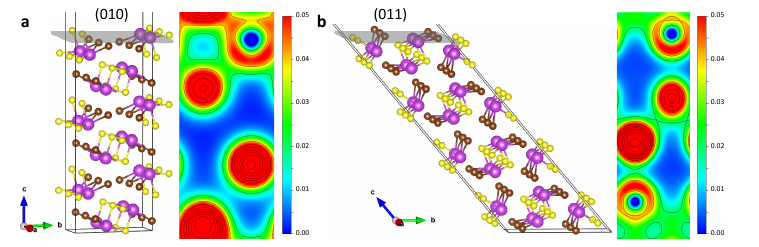}
        \caption{\textbf{Charge densities of relevant surfaces for BiSeI.}
        Relaxed slab structures and corresponding charge-density maps of the \textbf{a}~$(010)$ and \textbf{b}~$(011)$ surfaces. 
        The maps are evaluated on a plane parallel to the surface (grey plane in the structures), and highlight the different
        electronic distribution at the two facets.}
        \label{fig4}
\end{figure*}

Figure~\ref{fig3} reveals a systematic pattern: in all eight compounds, both band edges of the (011) surface lie at lower 
energies (i.e., deeper with respect to the vacuum level) than those of the (010) surface. The VBM of the (011) facet ranges 
from $-5.99$ to $-6.65~\mathrm{eV}$, whereas that of the (010) facet ranges from $-5.32$ to $-6.17~\mathrm{eV}$, and the 
CBM follows the same trend (Table~1). Since both facets belong to the same bulk crystal and therefore share the same band gap, 
the difference between them corresponds to a rigid shift of the entire band structure, with identical VBM and CBM displacements 
equal to the value $|\Delta V|$ listed in Table~1. This shift reflects a facet-dependent electrostatic potential step at the 
surface, consistent with the distinct electronic charge distributions of the two terminations (Fig.~\ref{fig4}), rather than 
any change in the bulk electronic structure. Provided that the Fermi level occupies the same position within the gap, $|\Delta V|$ 
also equals the work-function difference between the two facets (Fig.~\ref{fig1}c). Remarkably, $|\Delta V|$ lies between 
$0.63$ and $0.80~\mathrm{eV}$ in seven of the eight compounds ($\approx 0.7~\mathrm{eV}$ in average), which amounts to roughly 
35--55\% of the band gap and exceeds the thermal energy at room temperature ($k_\mathrm{B}T \approx 26~\mathrm{meV}$) by more 
than one order of magnitude.

The magnitude of the facet-induced shift depends only weakly on composition, with one notable exception. Replacing Bi by Sb 
moves the band edges of both facets upward by roughly $0.3$--$0.5~\mathrm{eV}$ on average, yet the resulting values of $|\Delta V|$ 
are comparable for Bi- and Sb-based compounds. The largest shifts, $0.80$ and $0.79~\mathrm{eV}$, are found for the sulfobromides 
BiSBr and SbSBr, and substituting I by Br in the sulfides increases $|\Delta V|$ by $0.11$--$0.16~\mathrm{eV}$. The clear outlier
is BiSeBr, with $|\Delta V| = 0.38~\mathrm{eV}$, about half the average of the rest of the family. In this compound, the band edges 
of the (010) surface lie $\approx 0.3~\mathrm{eV}$ lower than in BiSBr, whereas those of the (011) surface move in the opposite 
direction (Table~1). This observation suggests a termination-specific origin, such as a surface reconstruction, in line with the 
anomalous behaviour of BiSeBr already noted for the (021) surface energy (Fig.~\ref{fig2}a). Even in this case, however, the offset
remains far from negligible.

These results have two important implications. First, because the (011) and (010) facets are similarly stable, real films are 
likely to expose both of them, and any contact or transport layer can be optimised for only one of the two. An offset of 
$\sim 0.7~\mathrm{eV}$ between facets can be large enough to turn a selective interface on one facet into a recombination-prone, 
or extraction-blocking, interface on the other. Second, the pronounced dependence of the absolute band edges on surface orientation 
may contribute to the dispersion of the work functions and band edges reported experimentally for MChX films, since different 
synthesis conditions can favour the stabilisation of different facets. In the next section, we will quantify by means of device 
simulations how these facet-induced band-edge offsets affect photovoltaic performance.

Figure~\ref{fig4} shows the charge density of the (010) and (011) surfaces of BiSeI on a plane parallel to the surface, and reveals 
that the two terminations present markedly different electronic distributions. Although both maps display localised high-density 
regions associated with the outermost atoms, together with low-density regions in between, the number, position and lateral extent 
of these features differ substantially between the two facets, reflecting the different atomic arrangement and coordination 
environment exposed at each surface. These features are directly relevant to the band alignments of Fig.~\ref{fig3}. 

In our approach (Methods), the positions of the VBM and CBM relative to the average electrostatic potential are bulk quantities, so 
the facet dependence of the band edges (e.g., $-6.22$ and $-5.55$~eV for the VBM of the (011) and (010) surfaces of BiSeI, respectively; 
Table~1) enters entirely through the lineup of the electrostatic potential with the vacuum level. This lineup is set by the surface 
dipole, which is determined by how electronic charge is distributed and redistributed at the outermost atomic layers. Since the two 
facets have different charge distributions, they develop different potential steps across the surface, which shifts the whole band 
structure rigidly with respect to the vacuum level while leaving the band gap unchanged. The deeper band edges of the (011) surface 
are thus consistent with a stronger electron-attracting surface, in agreement with its higher electronegativity relative to (010), 
and the different charge distributions provide a natural microscopic origin for the $\sim 0.7$~eV facet-induced offset. Analogous 
charge-density differences are found for the rest of the MChX compounds, consistent with the systematic behaviour of Fig.~\ref{fig4}.

\begin{figure*}[t]
    \centering
        \includegraphics[width=1.0\textwidth]{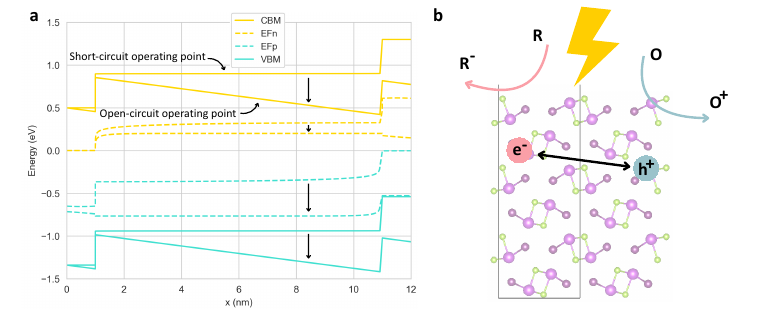}
        \caption{\textbf{Impact of MChX surface physics on photovoltaics.} 
        \textbf{a.}~Modeled energy band diagrams under illumination 
         for the Au/BiSBr~(010)/BiSBr~(bulk)/BiSBr~(011)/Au device at 1) short-circuit and 2) open-circuit operating points. The diagrams 
         reveal flat band conditions at $0$~V (short-circuit) and a progressively increasing band bending with forward bias (open-circuit). 
         Electron (E\textsubscript{Fn}) and hole (E\textsubscript{Fp}) quasi-Fermi levels confirm that the (010) ($x \to 0$~nm) facet 
         favors electron extraction, while the (011) ($x \to 12$~nm) facet enables hole collection. \textbf{b.}~Surface terminations 
         with suitably aligned VBM and CBM levels can drive also photocatalytic reactions.} 
        \label{fig5}
\end{figure*}

\subsection{Impact on photovoltaic technologies}
\label{subsec:photovoltaics}
Although the intrinsic optoelectronic properties of MChX suggest that they could deliver excellent photovoltaic performance, the 
power conversion efficiencies (PCEs) measured to date remain below 10\%, far from the Shockley--Queisser limit of $\sim$30\%
\cite{Shockley1961,Miliaieva2025}. Non-radiative recombination at point defects explains part of this gap~\cite{Lopez2025,Lopez2026},
but device-level limitations associated with surface-dependent band alignments are likely to be equally important
\cite{Klein2020,Colombara2020}. Photogenerated carriers must leave the absorber through its interfaces with the contact layers, and 
the band edges of the exposed facets determine whether these interfaces are selective or, on the contrary, open a direct recombination 
pathway. Since the (011) and (010) facets of MChX are close in energy and their band edges differ by up to $\approx 0.8$~eV 
(Fig.~\ref{fig3} and Table~1), we now examine how this asymmetry affects the performance of a photovoltaic device.

To quantify this effect, we performed drift-diffusion simulations with SCAPS-1D (Methods) \cite{scaps}, using BiSBr as a model system 
because it exhibits the most pronounced facet anisotropy: the work functions of its (010) and (011) surfaces differ by $\approx 0.8$~eV, 
the (011) facet being the more electronegative one. To isolate the role of this asymmetry, we considered a simplified 
Au/BiSBr(010)/BiSBr(bulk)/BiSBr(011)/Au device in which each facet is represented by a 1~nm thick layer with the electron affinity of 
the corresponding surface, the two contacts have the same work function of 5.0~eV (Au-like), and the thickness of the bulk layer, and 
hence the separation between the two facets, is varied (Methods, Table~2). All layers share the same transport and recombination 
parameters and no defect states are included, so that any change in device performance can be ascribed solely to the band-edge offset 
between the two facets.

The simulated current-voltage ($J$-$V$) curves reveal a steep decrease of the open-circuit voltage ($V_\mathrm{oc}$) as the separation 
between the (010) and (011) facets is reduced (Supplementary Fig.~9). For separations comparable to the grain sizes and carrier diffusion 
lengths expected in real absorbers, the voltage loss approaches $0.60$~V, roughly one third of the band gap of BiSBr. Figure~\ref{fig5}a 
shows the energy-band diagram of the device under illumination at the short-circuit and open-circuit operating points. The two facet 
layers give rise to steps in the band edges at the ends of the absorber, whose combined height amounts to the $\approx 0.8$~eV difference
between the two terminations. The bands are flat across the bulk layer at short circuit and progressively bend as the forward bias 
increases towards $V_\mathrm{oc}$. The quasi-Fermi levels ($E_{\mathrm{F}n}$ and $E_{\mathrm{F}p}$) show that the two facets play opposite 
roles: the (011) facet, with deeper band edges, favours electron extraction and repels holes, whereas the shallower (010) facet favours 
hole extraction and repels electrons. 

The facet asymmetry thus acts as an intrinsic, orientation-dependent selectivity. However, because the offset between facets is finite 
($\approx 0.8$~eV, less than half of the band gap), this selectivity is only partial. Carriers photogenerated within a diffusion length 
of a facet that is unfavourable for their extraction can still reach the contact of the wrong polarity and recombine there, which reduces 
the quasi-Fermi-level splitting and hence $V_\mathrm{oc}$. The shorter the separation between facets, the larger the fraction of carriers 
affected, in line with the trend observed in Supplementary Fig.~9.

These results show that, even when the absorber is chemically ideal and free of defects, the coexistence of facets with different band 
edges can suppress carrier extraction and limit device efficiency. A qualitatively similar facet-dependent photovoltaic response has
been reported for MAPb(I,Cl)$_3$ thin films~\cite{Gao2025}, which suggests that the effect may be general to other semiconductors. 
Conversely, the facet-dependent energetics can also be exploited: transport layers designed to match the band edges of the facets 
exposed at each interface could take advantage of the intrinsic driving force for charge separation provided by the facet offset, 
instead of counteracting it \cite{Li2020,Bai2015}. Device engineering through facet selection thus emerges as a powerful lever to 
improve efficiency, complementary to absorber optimisation through defect passivation and morphology control. We note that our 
one-dimensional model is a limiting case, as it places the two facets in series, whereas in real films they coexist laterally at 
the grain and interface level, so the quantitative voltage losses will depend on morphology.

\section{Discussion}
\label{sec:discussion}
Our results establish that surfaces play a decisive role in the optoelectronic functionality of pnictogen chalcohalides. 
Previous work has focused mainly on the lowest-index facets, such as (100), (010) and (001) \cite{Cano2025,Rovira2025,Lynch2026}. 
Our systematic exploration reveals that the previously overlooked (011) termination is not only thermodynamically competitive with 
(010), but also has markedly different band edges, shifted rigidly by $0.63$--$0.80$~eV in seven of the eight compounds studied
(Table~1). Since the two facets are close in energy, both may be stabilised in real samples depending on the synthesis conditions
\cite{synthesis1,synthesis2}. This mechanism offers a plausible explanation for the large variability in the work functions and 
band edges measured across MChX thin films, which could be tested with facet-resolved measurements on films of controlled texture.

The same facet-dependent energetics can also be exploited. Rather than being a loss mechanism, the offset between facets provides 
an intrinsic driving force for charge separation, which can be harnessed in devices whose transport layers are matched to the
band edges of the exposed facets \cite{Li2020,Bai2015}. This is particularly relevant for photocatalysis (Fig.~\ref{fig5}b). When 
facets with different VBM and CBM positions are exposed on the same particle, photogenerated electrons are driven towards the facet 
with the lower CBM and holes towards the facet with the higher VBM. This effect reduces recombination and makes both carriers 
available for reduction and oxidation reactions at spatially separated sites~\cite{Li2020,Bai2015}. BiSBr illustrates well this
possibility. Taking the H$^+$/H$_2$ and O$_2$/H$_2$O potentials at $-4.44$ and $-5.67$~eV relative to vacuum (pH = 0), the band 
edges of the (010) facet (VBM $=-5.85$~eV, CBM $=-4.01$~eV; Table~1) straddle both potentials, whereas those of the (011) facet 
(VBM $=-6.65$~eV, CBM $=-4.81$~eV) lie too low for proton reduction but provide strongly oxidising holes. Facet control could thus 
be used to tune both the thermodynamic driving forces and the spatial separation of carriers in MChX photocatalysts, although in 
a facet junction the electrons accumulate on the facet with the lower CBM, which may require cocatalysts or reactions other than 
hydrogen evolution.

Several caveats should be borne in mind. Our calculations consider clean, stoichiometric surfaces in vacuum, whereas real interfaces 
are influenced by reconstructions, adsorbates, non-stoichiometric terminations and surface states, which can modify both the absolute 
band edges and the facet offsets. In addition, the absolute positions of the band edges carry the uncertainty inherent to the level 
of theory, whereas the facet offset $|\Delta V|$ arises from the electrostatic lineup with the vacuum level and is expected to be 
more robust. The device model is likewise a limiting case that places the two facets in series, whereas in real films they coexist 
laterally at grain and interface level, so the actual voltage losses will depend on morphology. Experimental validation using 
Kelvin-probe force microscopy or photoemission spectroscopy on films of controlled texture, together with the extension of our approach 
to other terminations and to full device stacks, are natural next steps.

Together, these results highlight surface orientation as a hidden design parameter. Beyond the conventional focus on composition and 
defect chemistry, the control and stabilisation of specific crystal facets emerge as key levers for advancing MChX-based solar energy 
technologies.

\section{Conclusions}
\label{sec:conclusions}
We have presented a systematic first-principles study of the surface energetics and band-edge alignments of eight low-dimensional 
pnictogen chalcohalides (MChX; M = Bi, Sb; Ch = S, Se; X = I, Br). An exhaustive sampling of low- and higher-index orientations
identifies the (011) and (010) surfaces as the most stable facets in every compound, with formation energies of 
$0.08$--$0.14~\mathrm{J\,m^{-2}}$ that differ by only $0.013$--$0.034~\mathrm{J\,m^{-2}}$. The stability of the previously overlooked 
(011) facet shows that the magnitude of the Miller indices is a poor predictor of surface stability in these quasi-one-dimensional 
materials. Despite their similar energies, the two facets have markedly different band-edge alignments. The band edges of the (011) 
surface lie $0.63$--$0.80$~eV below those of the (010) surface in seven of the eight compounds ($0.38$~eV in BiSeBr), a rigid shift 
that we relate to the different electronic charge distributions, and hence surface dipoles, of the two terminations.

Drift-diffusion simulations of BiSBr show that the coexistence of facets with such offsets can reduce the open-circuit voltage by 
up to $0.6$~V, even in an otherwise ideal, defect-free absorber. Surface orientation therefore emerges as a hidden design parameter
that, together with non-radiative recombination at point defects, may contribute to the gap between the experimental and theoretical 
efficiencies of MChX solar cells. Conversely, the same offsets provide an intrinsic driving force for charge separation that can be 
exploited in photovoltaic and photocatalytic devices by matching the contact and transport layers to the exposed facets. These conclusions 
are based on clean, stoichiometric surfaces and on a one-dimensional device model, so experimental validation on films of controlled 
texture, for instance by Kelvin-probe force microscopy or photoemission spectroscopy, is a natural next step. More broadly, the workflow 
presented here, which links atomistic surface energetics to device-level performance, is transferable to other emerging semiconductors in
which facet selection may be equally decisive. Explicitly accounting for surface orientation is thus a key step towards unlocking the 
full potential of MChX in solar energy conversion.

\section*{Methods}
\label{sec:methods}
\textbf{First-principles calculations.}~\textit{Ab initio} calculations based on density functional theory (DFT) were  performed to 
analyse the physicochemical properties of bulk pnictogen chalcohalides. These calculations were conducted with the \verb!VASP! software 
package \cite{Kresse1996} using the generalized gradient approximation to the exchange-correlation energy for solids due to Perdew 
\textit{et al.} (PBEsol) \cite{Perdew2008}. Spin–orbit coupling effects, which are particularly relevant for Bi-based MChX  
\cite{Schimka2011,Heyd2003,Krukau2006}). The projector augmented-wave method was used to represent the ionic cores \cite{Blochl1994} 
and for each element the maximum possible number of valence electronic states was considered. Wave functions were represented in a 
plane-wave basis typically truncated at $450$~eV. By using these parameters and dense $\Gamma$-centered \textbf{k}-point grids for 
reciprocal space Brillouin zone integration, the resulting energies were converged to within $1$~meV per formula unit. In the geometry 
relaxations, a tolerance of $0.5$~meV$\cdot$\AA$^{-1}$ was imposed in the atomic forces. In every calculation, a $|\textbf{a} \times 
\textbf{b}| \geq 40$ density of \textbf{k}-points was always maintained for any bulk and surface calculation.
\\

\begin{table*}[t]
    \centering
    \begin{tabular}{cccccc}
    \hline
    \hline
        Material                                                      & Back Contact & BiSBr (010) & BiSBr Bulk            & BiSBr (011) & Front Contact \\ \hline
        Thickness (nm)                                                & -            & 1           & Variable              & 1           & -             \\
        Band Gap (eV)                                                 & -            & 1.84        & 1.84                  & 1.84        & -             \\
        Electron Affinity (eV)                                        & 5.0          & 3.7         & Graded $\beta$-function & 4.5         & 5.0           \\
        Dielectric Permitivity (relative)                             & -            & 10          & 10                    & 10          & -             \\
        CB effective density of states (cm\textsuperscript{3})        & -            & $10^{19}$   & $10^{19}$             & $10^{19}$   & -             \\
        VB effective density of states (cm\textsuperscript{3})        & -            & $10^{19}$   & $10^{19}$             & $10^{19}$   & -             \\
        Electron thermal velocity (cm/s)                              & -            & $10^{7}$    & $10^{7}$              & $10^{7}$    & -             \\
        Hole thermal velocity (cm/s)                                  & -            & $10^{7}$    & $10^{7}$              & $10^{7}$    & -             \\
        Electron Mobility (cm\textsuperscript{2}/Vs)                  & -            & 50          & 50                    & 50          & -             \\
        Hole Mobility (cm\textsuperscript{2}/Vs)                      & -            & 50          & 50                    & 50          & -             \\
        Shallow Uniform donor density (cm\textsuperscript{-3})        & -            & $10^{5}$    & $10^{5}$              & $10^{5}$    & -             \\
        Shallow Uniform acceptor density (cm\textsuperscript{-3})     & -            & $10^{5}$    & $10^{5}$              & $10^{5}$    & -             \\
        Absorption Coefficient (eV\textsuperscript{1/2}/cm)           & -            & $10^{5}$    & $10^{5}$              & $10^{5}$    & -             \\
        Radiative Recombination coefficient (cm\textsuperscript{3}/s) & -            & $10^{-10}$  & $10^{-10}$            & $10^{-10}$  & -
    \\ 
    \hline
    \hline     
    \end{tabular}
    \caption{\textbf{Details of the SCAPS-1D simulations.} The bulk BiSBr layer features a graded electron affinity implemented via a 
             $\beta$-function to model a smooth transition between the (010) and (011) facets. All layers share identical transport and 
             recombination parameters, and contact work functions were fixed at $5.0$~eV.}
    \label{table2}
\end{table*}

\textbf{Band alignment calculations.}~To estimate accurate band alignments, we followed the first-principles computational strategy
employed in previous works \cite{Liu2020,Liu2022}. Briefly, both bulk and slab calculations were performed from which the alignment
of the electrostatic potential within the dielectric material can be obtained relative to the vacuum level. From the slab calculations,
the average electrostatic potential of the material relative to the vacuum level was obtained. From the bulk calculations, the valence
band maximum and conduction band minimum relative to the average electrostatic potential were determined. These calculations involved
the estimation of macroscopic and planar average potentials. The planar potential was computed by averaging potential values within
a well-defined plane (for instance, perpendicular to the surface of the slab), and the macroscopic potential was obtained by taking
averages of the planar potential over distances of one unit cell along the chosen direction. The slab systems should be thick enough
to ensure that the electron density in the center of the slab is practically equal to that in the bulk material. We found that
$\approx 15$~\AA~ thick slabs accompanied by similarly large portions of vacuum provided sufficiently well converged electrostatic
potentials.
\\

\textbf{Optical solar cell device modeling.}~Drift-diffusion simulations were performed using SCAPS-1D \cite{scaps}. The full set of 
simulation parameters is summarized in Table~2, and all SCAPS setting and output files are provided in the Supporting Information to 
ensure reproducibility. To model the facet-dependent work function variation, the electron affinity and doping levels were adjusted to 
yield an effective work function difference across the junction. A smooth spatial transition between facet properties was implemented 
using a beta-function profile, enabling a constant and averaged work function in the intermediate layer.
\\

\section*{Data availability}
The data that support the findings of this study have been made publicly available \cite{database}, comprising the geometrical relaxations 
and local potential calculations for all relaxed slabs.
\\

\section*{Acknowledgments}
C.C. acknowledges support by MICIN/AEI/10.13039/501100011033 and ERDF/EU under the grants CNS2025-165467,
PID2023-146623NB-I00 and PID2023-147469NB-C21 and by the Generalitat de Catalunya under the grants 2021SGR-00343,
2021SGR-01519 and 2021SGR-01411. Computational support was provided by the Red Española de Supercomputación under
the grants FI-2024-1-0005, FI-2024-2-0003, FI-2024-3-0004, FI-2024-1-0025, FI-2024-2-0006, and FI-2025-1-0015.
This work is part of the Maria de Maeztu Units of Excellence Programme CEX2023-001300-M funded by MCIN/AEI
(10.13039/501100011033). E.S. acknowledges the European Union H2020 Framework Program SENSATE project: Low dimensional
semiconductors for optically tuneable solar harvesters (grant agreement Number 866018), Renew-PV European COST
action (CA21148) and the Spanish Ministry of Science and Innovation ACT-FAST (PCI2023-145971-2). E.S. is grateful
to the ICREA Academia program.
\\

\section*{Author contributions}
C.L. and C.C. conceived the study and planned the research, which was discussed in-depth with the rest of the co-authors. C.L. and C.C. 
performed and analyzed the first-principles calculations. C.L. carried out the data analysis of the generated DFT computations. Z.J. and 
A.J. performed the simulations of multi-junction solar cell devices. The manuscript was written by C.L. and C.C. with substantial input 
from the rest of the co-authors.
\\

\section*{Additional information}
Supplementary information is available in the online version of the paper.
\\

\section*{Competing financial interests}
The authors declare no competing financial interests.
\\

\bibliography{Bibliography}

\end{document}